\documentclass[a4paper, amsfonts, amssymb, amsmath, reprint, showkeys, nofootinbib, twoside]{revtex4-1}
\usepackage[english]{babel}
\usepackage[utf8]{inputenc}
\usepackage[colorinlistoftodos, color=green!40, prependcaption]{todonotes}
\usepackage{amsthm}
\usepackage{amsmath,amssymb,amsfonts}
\usepackage{mathtools}
\usepackage{physics}
\usepackage{xcolor}
\usepackage{graphicx}
\usepackage[left=23mm,right=13mm,top=35mm,columnsep=15pt]{geometry}
\usepackage{adjustbox}
\usepackage{placeins}
\usepackage[T1]{fontenc}
\usepackage{lipsum}
\usepackage{csquotes}
\usepackage[normalem]{ulem}

\begin{document}
\title{Penetrating a granular medium by successive impacts}

\author{Antoine Seguin}
\affiliation{Universit\'e Paris-Saclay, CNRS, Laboratoire FAST, 91405 Orsay, France}

\author{Yann Bertho}
\affiliation{Universit\'e Paris-Saclay, CNRS, Laboratoire FAST, 91405 Orsay, France}

\author{Baptiste Darbois Texier}
\affiliation{Universit\'e Paris-Saclay, CNRS, Laboratoire FAST, 91405 Orsay, France}

\begin{abstract}
We consider the penetration dynamics of a vertical cylinder into a dry granular medium subjected to successive impacts. The depth of the impactor below the free surface $z_N$ first evolves linearly with the impact number $N$ and then follows a power-law evolution $z_N \propto N^{1/3}$. The depth reached by the cylinder after a given number of impacts is observed to increase with the impact energy but to decrease with its diameter and the density of the granular medium. We develop a model that accounts for the quasi-static and inertial granular forces applying on the cylinder to rationalize our observations. This approach reveals the existence of two intrusion regimes for large and small impact numbers, allowing all data to be rescaled on a master curve. Then, we extend the study to the effect of sidewalls on the dynamics of the impactor. We show that lateral confinement changes the dependence of the impactor depth on the impact number $z_N (N)$. This effect is accounted for by considering the increase of the granular drag with the lateral confinement.
\end{abstract}

\maketitle

\section{Introduction}
\label{intro}
Probing the ground to extract samples or to dig up buried objects is at the heart of many civil engineering activities and exploration missions. Several applications, such as tunneling or underground scanning, would benefit from the development of subterranean robots capable of penetrating and moving below the surface \cite{Aguilar2016,naclerio2021controlling}. Among the techniques used to penetrate a probe into the ground, several rely on the transfer of kinetic energy to the digging impactor as in drilling~\cite{Shi2019}. This transfer can also be done using a succession of steps like when someone drives a nail, i.e., hammering. In particular, the hammering mechanism underpins the operation of space probes that are designed to dive below the sandy surface of other planets and inspect their physical properties~\cite{Spohn2018,Olaf2019}. Despite its common use, the hammering technique as a means of penetrating a granular soil is little explored from a physical point of view.

The study of granular media has followed an intensive effort these last decades, with a particular focus on the stress state within these materials. The pioneering work in this field is by Janssen who considered the stress distribution in a confined granular medium~\cite{Janssen1895}. He observed that the pressure exerted on the base of the container saturates at a value much smaller than the hydrostatic pressure that would be measured for a liquid column. Instead, the pressure saturates at a constant value that scales as the weight of grains over a vertical distance comparable to the width of the container. His continuum phenomenological model relies on the hypothesis that the frictional contact forces between the grains and the walls are at their Coulomb threshold and that the vertical stresses are redistributed proportionally to the horizontal ones. This effect, characteristic of granular media, has led to numerous works in static \cite{Ovarlez2003,Cambau2013,Mahajan2020} or dynamical configurations \cite{Bertho2003, Bertho2004}.

Over the last few decades, many works were devoted to the dynamics of an object impacting a granular medium. A large part of this effort concerns spherical projectiles, well suited to the study of impact crater morphology \cite{Debruyn2004penetration,Deboeuf2009,Van2017,Huang2020}. This configuration has implications in geophysics for meteorite impacts \cite{Melosh1989,Katsuragi2016}, a problem in which the effective gravity must be taken into account \cite{Altshuler2014}. Different experiments and numerical simulations permitted to estimate the drag force applying on the sphere during its intrusion \cite{Uehara2003,Ambroso2005,Katsuragi2007,Goldman2008,Seguin2008,Seguin2009,Umbanhowar2010,Pacheco2011,Nordstrom2014,Joubaud2014}. In parallel, theoretical models were proposed to rationalize these observations \cite{Tsimring2005,Hinch2014,Guo2018}. A widely used model to describe the drag force during the impact of a sphere is that which considers the sum of two contributions: a collisional term and a Coulomb friction term \cite{Katsuragi2007}. The collisional term evolves as $~v^2$, where $v$ is the penetration velocity of the sphere, and is associated to the energy dissipation by collisions with the grains constituting the granular medium \cite{Takehara2010,Takehara2014}. This expression for the inertial drag was shown to hold for objects of different shapes with a prefactor that accounts for their geometry \cite{cheng2018collision}. The frictional term, by analogy with the hydrostatic pressure in a classical fluid, is described with an evolution as $\rho g z$, where $z$ is the penetration depth of the object \cite{kang2018archimedes}. This prediction holds for small depths and different models were proposed for the drag force at larger depths. In this limit, the quasi-static drag force on a penetrating object was described by a generic term in $z^{\alpha}$, with $\alpha>1$ and that depends on the object geometry \cite{Hill2005,Peng2009,Brzinski2013}. More recently, the quasi-static granular drag experienced by a slender object was rationalized as the sum of a $z$ term and a $z^2$ term \cite{Seguin2018}. This model is the result of two contributions: the pressure of the granular force at the tip of the slender object ($\propto z$) and the granular frictional force on the lateral side of the submerged part, integrated over the entire depth ($\propto z^2$). The previous expressions for the drag force permitted to capture finely the dynamics of an object that realizes a single impact in a granular medium. However, the predictions of these models regarding a succession of impacts remain to be explored.

In this paper, we study experimentally the penetration of a slender object (called impactor hereafter) into a dry granular material by successive impacts. First, we measure the evolution of the penetration depth of the impactor as a function of the number of impacts it receives, in a granular medium that can be considered as infinite. We highlight the existence of two regimes and we develop a continuous modeling of the penetration depth. Second, we explore the influence of a lateral confinement during the penetration and we show that friction to the walls introduces a modification of the scaling law of the penetration depth.

\section{Experimental setup}
\label{setup}
\begin{figure}[t]
\centering
\includegraphics[width=\hsize]{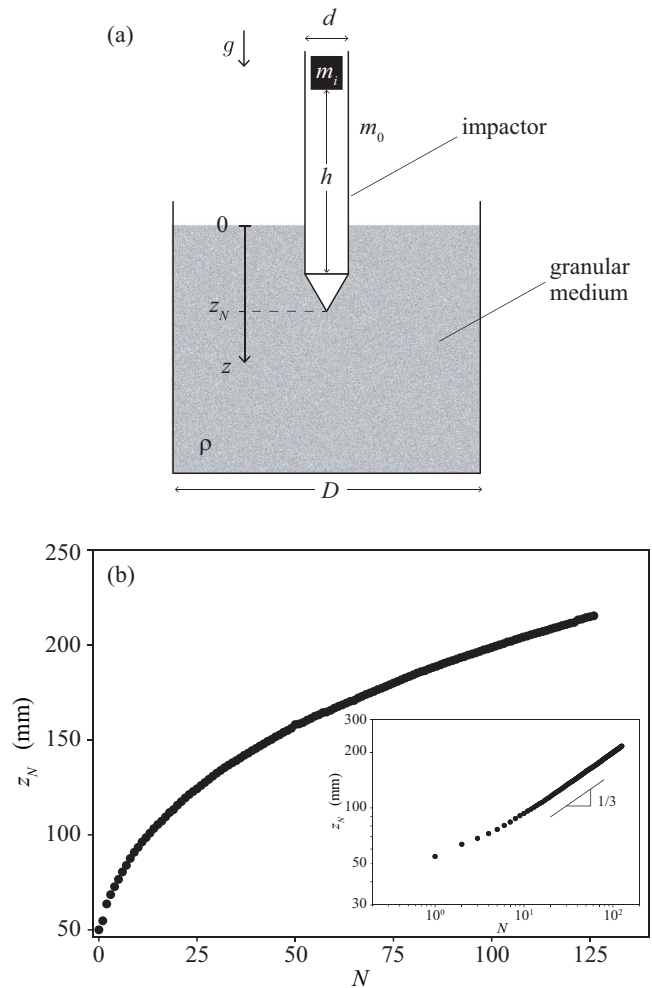}
\caption{(a) Sketch of the experimental setup and notations used. (b) Depth of the impactor $z_N$ as a function of the impact number $N$, for an intrusion in glass beads ($d_g=0.45$~mm and $\rho=1550$~kg~m$^{-3}$). The impactor has a diameter $d=50$~mm and a mass $m_0=460$~g. The impacting mass is $m_i=209$~g and is released from $h=25.5$~cm corresponding to an energy $\mathcal{E}=0.52$~J. Inset: Same data in a logarithmic plot. The black line represents $z_N \propto N^{1/3}$.}
\label{Fig01}
\end{figure}

The experimental setup consists of a cylindrical container of diameter $D$, height 300~mm, containing rigid beads of diameter $d_g$ and density $\rho_g$ [Fig.~\ref{Fig01}(a)]. The beads used are made of different materials, plastic, glass, and metal of density $\rho_g=920$, 2500, and 7740~kg~m$^{-3}$, and diameters $d_g=1$, 0.45 and 2~mm, respectively. A controlled preparation of the granular medium is obtained by pouring the grains into the container and then tapping it at least ten times on the floor, ensuring an initial volume fraction $\phi=0.62 \pm 0.01$. We define the bulk density of the granular medium as $\rho = \phi \, \rho_g$. The impactor is a hollow cylinder of external diameter $d$ (ranging from 16 to 50~mm) and mass $m_0$ (ranging from 60 to 460~g) ending with a $60^\circ$ angle conical head, as sketched in Fig.~\ref{Fig01}(a). Initially, the tip of the impactor is placed at a depth $z_0 = 50$~mm below the surface of the granular medium to ensure its stability. The impacts are generated by a cylindrical mass $m_i=209$~g dropped inside the hollow cylinder from a height $h$, without any initial velocity. When the mass reaches the bottom of the impactor, it transfers its momentum and the impactor moves deeper in the granular medium. The impacts are reproduced several times in a row, keeping constant the dropping height $h$ and thus the impact energy $\mathcal{E}=m_i gh$. A picture of the impactor is taken after each impact and allows one to measure its final depth $z_N$ below the granular surface as a function of the impact number $N$. Particular care has been taken to systematically center the impacting mass with the axis of the impactor to ensure that the system remains vertical during successive impacts (deviation less than $2^\circ$). In addition, we checked that the time interval between impacts has no influence on the intrusion dynamics in a range between 10~s and 100~s. This interval time is much larger than the characteristic time required for a grain to fall from its own diameter $\sqrt{d_g/ g} \simeq 10$~ms. In all the experiments, the ratio of the impactor diameter over the bead diameter $d/d_g$ is always larger than 10, allowing to describe the medium in the limit of continuum mechanics \cite{Seguin2018}. Note that we do not measure significant variations of the mean packing fraction during the successive impacts, as each impact reinforces the initial tapping protocol. Also, to avoid any interaction with the bottom of the container, the distance between the tip of the impactor and the bottom is at least ten times larger than the diameter of the grains \cite{stone2004local}.

\begin{figure*}[t]
\centering
\includegraphics[width=\hsize]{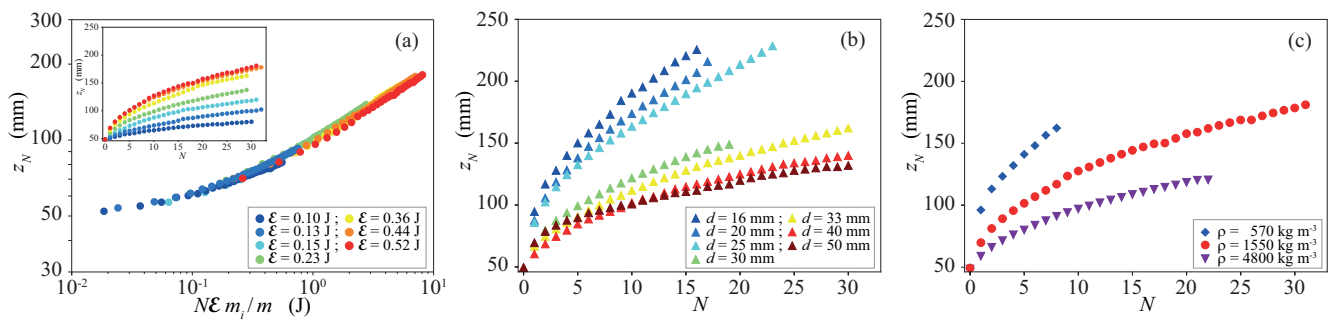}
\caption{(a)~Depth of the impactor $z_N$ as a function of the accumulated impacting energy $N \mathcal{E}\, m_i/m$ for different impact energies $\mathcal{E}$. Inset: $z_N$ as a function of the impact number $N$. For intrusions in glass beads ($d_g=0.45$~mm and $\rho=1550$~kg~m$^{-3}$), the impactor has a diameter $d=33$~mm, a mass $m_0=210$~g and the impacting mass is $m_i=209$~g. (b)~Depth $z_N$ as a function of the impact number $N$ for different diameters $d$, with $d_g=0.45$~mm and $\rho=1550$~kg~m$^{-3}$, by keeping constant the impact energy $\mathcal{E}$ and the mass ratio $m_i/ m$. (c)~Depth $z_N$ as a function of the impact number $N$ for different densities of the granular medium $\rho$, with $d=33$~mm and a mass $m_0=210$~g, the impacting mass $m_i=209$~g is released from $h=255$~mm.}
\label{Fig02}
\end{figure*}
\section{Penetration in unbounded granular medium}
\subsection{Experimental results}
\label{results}
We first consider the case where the diameter of the container is large compared to the impactor diameter ($D>8\,d$) and side walls have a negligible effect on the impactor penetration~\cite{Seguin2008}. A typical penetration dynamics obtained under these conditions is shown in Fig.~\ref{Fig01}(b) for an impactor entering into glass beads and experiencing $N=125$ successive impacts. We observe that the dynamics is fast for the first impacts then slows down with the number of impacts, but continues to progress without saturating with depth. The inset in Fig.~\ref{Fig01}(b) displays the logarithmic plot of the data and shows that the penetration dynamics is well approximated by the scaling law $z_N \propto N^{1/3}$ when $N \gtrsim 10$.

The inset in Fig.~\ref{Fig02}(a) presents the penetration dynamics $z_N$ for an impactor entering glass beads and submitted to different impact energies $\mathcal{E}$. We observe that the intrusion dynamics keeps the same increasing shape whatever the impact energy, but that the depth after a given number of impacts increases with $\mathcal{E}$. The same data are presented in Fig.~\ref{Fig02}(a) as a function of the total impact energy delivered to the impactor $N \mathcal{E}\, m_i/m$, where $m=m_0 + m_i$. It turns out that experiments performed with different pairs of $h$ and $m_i$ but the same impact energy $\mathcal{E}= m_igh$ leads to the same results, highlighting that the impact energy is the relevant parameter in this problem.

The effect of the impactor diameter $d$ on the penetration depth is displayed in Fig.~\ref{Fig02}(b) as a function of the impact number $N$, by keeping constant the impact energy $\mathcal{E}$. The intrusion dynamics is then observed to be slower for larger impactor diameters.

Finally, we explore the influence of the density of the granular medium on the penetration by successive impacts [Fig.~\ref{Fig02}(c)]. We realize experiments with grains of different materials by keeping unchanged the impactor characteristics ($m_0=210$~g, $m_i=209$~g, and $d=33$~mm) as well as the impact energy ($\mathcal{E}=0.52$~J). We observe that increasing the bulk density $\rho$ of the granular medium significantly slows down the intrusion dynamics.

\subsection{Model}
To model the evolution of the penetration depth $z_N$ with the number $N$ of successive impacts, we consider the forces that apply on the impactor during the advancing phases. The impactor experiences its own weight, the force resulting from the impact of the internal mass, and the resistance of the granular medium. Here, the weight of the impactor $m g$ is small compared to the granular drag and can be neglected in the force balance. The mechanical action exerted by the impacting mass is accounted through momentum conservation during the collision of the internal mass with the bottom of the cylinder. The impacting mass $m_i$ dropped from a height $h$ without initial velocity has a mechanical energy $m_igh$ and a velocity before impact $\sqrt{2gh}$. The collision of the internal mass $m_i$ with the penetrating object of mass $m_0$ is assumed to be perfectly inelastic. In these conditions, the conservation of the momentum of the system provides that the initial velocity of the impactor just after the impact is $v_i=\sqrt{2gh}\, m_i/m$. The resistive forces resulting from the interaction of the impactor with the granular material have both a collisional and a frictional origin. At each new impact, two temporal regimes can be distinguished \cite{Seguin2009}: a first phase corresponding to the earliest instants where the inertial resistance dominates and where the drag force strongly depends on the velocity of the impactor; a second phase where the quasi-static resistance dominates, independent of the impactor velocity. In the following, we will focus on the terms corresponding to these two regimes, one after the other.

By considering first the case where the velocity of the impactor is high and the inertial contribution of granular drag, proportional to the cross sectional area ($\sim d^2$) and to the density of the granular medium $\rho$, overcomes the quasi-static contribution, the equation of motion for the impactor of mass $m$ is written as
\begin{equation}
m\frac{dv}{dt}= - K_v\, \rho d^2\, v^2,
\label{eq1}
\end{equation}
where $v=dz/dt$ and $K_v$ is a characteristic prefactor of the collision dissipation. As in the case of a sphere impact in a granular medium, $K_v$ is expected to depend on the packing fraction $\phi$ \cite{Debruyn2004penetration,Seguin2009,Guo2018}. The model is developed in the limit of continuous depths $z$ and then discretized to $z_N$ to be compared with experiments. At the beginning of impact number $N$, the impactor is located at the depth $z=z_{N-1}$ of the previous impact and its velocity is $v=v_i$. By using the relation $dv/dt=(dv /dz)(dz/dt)=v\, dv/dz=(1/2)\,d(v^2)/dz$, Eq.~(\ref{eq1}) leads to
\begin{equation}
\frac{v^2(z)}{v_i^2}=\exp\left(-\frac{z-z_{N-1}}{L_0}\right),
\end{equation}
where $L_0=m/2K_v\, \rho d^2$ is a characteristic length, which scales as $m/\rho \, d^2$.
This expression is valid at the first instants of the penetration dynamics. The final depth $z_{N}$ is reached when the impactor velocity vanishes, i.e., $v(z_N)=0$. We obtain a relation between two successive rest positions $z_N-z_{N-1}=L_0$, which, by iteration from the initial depth $z_0$, leads to the following prediction:
\begin{equation}
\frac{z_N-z_0}{L_0} = N.
\label{smallz}
\end{equation}
This relationship describes the first asymptotic regime that is expected to occur during the first set of impacts at shallow depths.

We now consider the opposite situation where the quasi-static contribution of the drag force overcomes the inertial term and which occurs in the limit of low velocities or high depths. In this limit, the resistance experienced by the impactor from the granular medium is proportional to $z^2$ \cite{Albert1999,Seguin2018}. We neglect here the resistive force term of the pressure at the tip of the impactor, which should evolve as $z$, assuming it is approximately cylindrical. Under these conditions, the equation of motion for the penetration dynamics of the impactor is written as
\begin{equation}
m\frac{dv}{dt}= - K_z\, \rho gd\, z^2,
\end{equation}
where $K_z$ is a characteristic prefactor of the friction force. This equation can be made non-dimensional by using $L_0$ and by introducing a characteristic time $T_0=(K_v\, d/K_z\, g)^{1/2}$, which scales as $(d/g)^{1/2}$. For the impact number $N$, the initial conditions are $z=z_{N-1}$ and $v=v_i$. Using the same relation as before for $dv/dt$, we obtain the following solution:
\begin{equation}
\frac{v^2(z)}{v_i^2}= 1- \frac{1}{3}\left(\frac{L_0}{v_i T_0}\right)^2 \left(\frac{z^3-z_{N-1}^3}{L_0^3}\right).
\end{equation}
The final depth $z_{N}$ is reached when $v(z_N)=0$. We obtain a relation between two successive impacts, which, by iteration with respect to the initial depth, leads to the following prediction:
\begin{equation}
    \frac{z_N^3 - z_0 ^3}{L_0 ^3} = 3 N \left(\frac{T_0 \, v_i}{L_0} \right)^2.
    \label{bigz}
\end{equation}
In the limit of large depths ($z_N \gg z_0$), we can approximate that $z_N^3 - z_0 ^3 \simeq (z_N - z_0)^3$ and the previous equation leads to:
\begin{equation}
    \frac{z_N - z_0}{L_0} = \left[3N \left(\frac{T_0 \, v_i}{L_0} \right)^2 \right]^{1/3}.
    \label{bigzapprox}
\end{equation}
\begin{figure*}[t]
\centering
\includegraphics[width=\hsize]{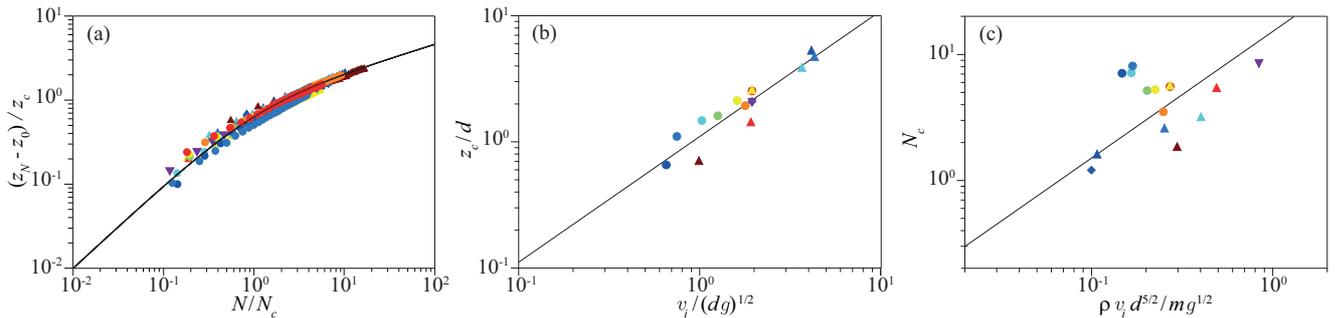}
\caption{(a)~Normalized impactor depth $(z_N - z_0)/z_c$ as a function of the normalized impact number $N/N_c$. The solid line represents the prediction of Eq.~(\ref{eq:merge}). (b)~Crossover distance $z_c$ deduced from the fit and normalized by the impactor diameter $d$ as function of the experimental parameter $v_i/(d\, g)^{1/2}$. The solid line corresponds to $z_c/d= A\, v_i/(d\, g)^{1/2}$, with $A=1.1$. (c)~Crossover impact number $N_c$ as a function of the experimental parameter $\rho v_i\, d^{5/2}/(m\, g^{1/2})$. The solid line corresponds to $N_c=B\rho v_i\, d^{5/2}/(m\, g^{1/2})$, with $B=15$. Symbols have the same meaning as in Fig.~\ref{Fig02}.}
\label{fig:crossover}
\end{figure*}
\indent To summarize, two intrusion regimes should be considered, where $z_N\sim m N/\rho d^2 \propto N$ in the limit of small depths, and $z_N\sim (m v_i^2 N / \rho g d)^{1/3} \propto N^{1/3}$ in the limit of large depths. Note that both predictions provide that the depth of the impactor after a given number of impacts decreases with the impactor diameter $d$ and the density of the granular medium $\rho$, in qualitative agreement with experiments displayed in Figs.~\ref{Fig02}(b) and \ref{Fig02}(c). At intermediate depths, we expect a crossover between these two regimes and we suggest to connect these two predictions by the arbitrary expression
\begin{equation}
    \displaystyle \frac{z_N - z_0}{z_c}=\frac{ \displaystyle \frac{N}{N_c}}{ \displaystyle \left(1+\frac{N}{N_c}\right)^{2/3}},
    \label{eq:merge}
\end{equation}
where $z_c$ is the crossover depth and $N_c$ the number of impacts associated to this crossover. Note that shallow penetration depths corresponding to $N/N_c \ll 1$ and large penetration depths corresponding to $N/N_c \gg 1$, lead, respectively, to the limiting cases $z_N \propto N$ and $z_N \propto N^{1/3}$ discussed before. We consider $z_c$ and $N_c$ as free parameters and we seek the best fit of the measured intrusion dynamics with Eq.~(\ref{eq:merge}). Figure~\ref{fig:crossover}(a) presents all the intrusion dynamics normalized by $N_c$ and $z_c$ estimated through this procedure. We conclude that the proposed modeling is compatible with the experimental data since all our measurements collapse on a master curve which corresponds to Eq.~(\ref{eq:merge}). The identification of Eqs.~(\ref{smallz}), (\ref{bigz}), and (\ref{eq:merge}) leads to the following expressions for the crossover parameters: $z_c =\sqrt{3} \, v_i \, T_0 \sim v_i (d/g)^{1/2}$ and $N_c = \sqrt{3}\, T_0 \, v_i /L_0 \sim \rho v_i\, d^{5/2} / (m \, g^{1/2}) $. In the following, we compare the estimations provided by our fitting procedure to these predictions. Figure \ref{fig:crossover}(b) shows the evolution of $z_c /d$ as a function of $v_i/(d\, g)^{1/2}$. We observe a good agreement between the experimental data and the scaling of $z_c$ predicted by this modeling. The fit of the data highlighted by the solid line in Fig.~\ref{fig:crossover}(b) provides an estimate of the ratio $K_v/K_z=0.4 \pm 0.2$. Figure~\ref{fig:crossover}(c) presents the evolution of $N_c$ as a function of $\rho v_i \, d^{5/2} / (m\, g^{1/2})$. The fit of the data [solid line in Fig.~\ref{fig:crossover}(c)] provides an estimate of the ratio $({K_v}^3/K_z)^{1/2}=4 \pm 3$. The agreement between the measurements and predictions is still reasonable even though there is more dispersion on the data. This dispersion arises from the fact that $N_c$ is limited to a narrow range of values (between 1 and 10) while the uncertainty on this parameter can hardly be less than 1. Moreover the variations of prefactors in granular media are known to be highly fluctuating \cite{kang2018archimedes,Seguin2018}. These two scaling laws allow to give an estimate of $K_v\simeq 6 $ and $K_z\simeq 15$. Note that the values of $K_v$ and $K_z$ are known to depend on the shape of the impactor. Indeed, a collisional model permitted to propose a dependency for $K_v$ with the geometry of the object \cite{cheng2018collision}. In the quasi-static limit, an Archimedean law was proposed for the granular drag at low depths, which permits to express the influence of the head geometry on the resistive force \cite{kang2018archimedes}.

The crossover between the two intrusion regimes is described by the ratio between the collisional and the frictional forces. The transition between these two regimes can be rationalized with a non-dimensional number which corresponds in this case to the Bagnold number $\mathrm{Ba}$ defined as:
\begin{equation}
\mathrm{Ba}=\frac{\rho d^2 v^2}{\rho gd z^2}= \left(\frac{d}{z}\right) \left(\frac{v^2}{gz}\right)=\left(\frac{d}{z}\right) \mathrm{Fr}^2,
\end{equation}
where $\mathrm{Fr}=v/\sqrt{gz}$ is a dynamic Froude number. This definition drops the constant relative to the ratio $K_v / K_z$ which is of the order of unity. When $\mathrm{Ba} \ll 1$, the quasi-static contribution of the granular resistance predominates, whereas when $\mathrm{Ba} \gg 1$, the inertial term overcomes. In these experiments, this number is related to the Froude number which compares two characteristic velocities: the initial velocity due to the impact of the mass and the velocity linked to the hydrostatic pressure. The dynamic side comes from the fact that the more the object sinks, the smaller this number naturally becomes.

\section{Penetration in a confined granular medium}
\label{Discuss}
\begin{figure}[t]
\centering
\includegraphics[width=\columnwidth]{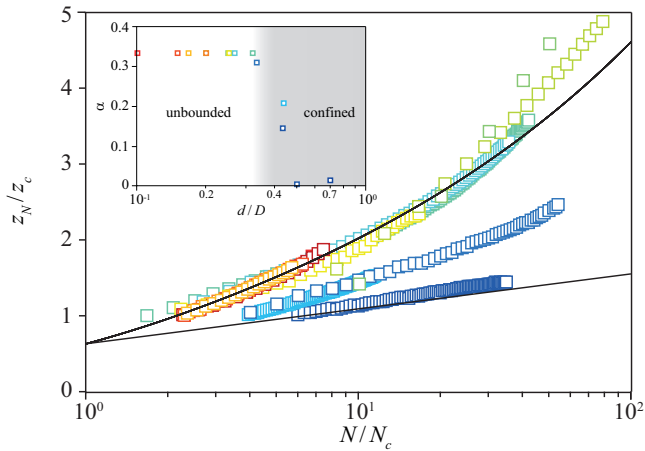}
\caption{Normalized depth $z_N/z_c$ as a function of the normalized impact number $N/N_c$ for increasing ratio of impactor diameter over the reservoir diameter $d/D$. The upper solid line indicates the prediction for unbounded intrusion from Eq.~(\ref{eq:merge}) and the lower solid line represents the extreme case $z_N \propto \ln \, N$ when the lateral confinement matters. Inset: Estimated exponent $\alpha$ from the fit of the intrusion dynamics with Eq.~(\ref{eq:merge_alpha}) as a function of $d/D$. The gray region distinguishes confined cases from the unbounded case.}
\label{fig:zj}
\end{figure}

In this section, we consider the effect of a lateral confinement on the intrusion dynamics of the impactor over successive impacts. We follow the same experimental procedure as previously, but we now investigate larger ratio of the impactor diameter over reservoir diameter $d/D$ where the influence of the lateral walls is not negligible anymore. Whereas previous experiments were conducted for $d/D<0.125$, we now increase this ratio up to $0.4$. Figure~\ref{fig:zj} shows the intrusion dynamics of an impactor penetrating into glass beads with different values of the ratio $d/D$. When $d/D$ is smaller than 0.3, all the intrusion dynamics collapse on the prediction of Eq.~(\ref{eq:merge}) when rescaled by the crossover depth $z_c$ and the impact number $N_c$. However, when $d/D$ becomes larger than 0.3, the intrusion dynamics starts to deviate from the unbounded case. The lateral confinement not only reduces the depth of the impactor after a given number of impacts, but also changes the penetration law as highlighted by the semi-logarithmic scales of Fig.~\ref{fig:zj} that reveal a transition towards a logarithmic trend, $z_N \propto \ln~N$, when $d/D \gtrsim 0.3$. To characterize this change in behavior, we fit the intrusion dynamics by the following law:
\begin{equation}
    \displaystyle \frac{z_N - z_0}{z_c}=\frac{ \displaystyle \left(\frac{N}{N_c}\right)^{ \alpha + 2/3}}{ \displaystyle \left(1+\frac{N}{N_c}\right)^{2/3}},
    \label{eq:merge_alpha}
\end{equation}
where $\alpha$ is a free parameter equals to $\alpha=1/3$ when the effect of the lateral confinement is negligible, recovering Eq.~(\ref{eq:merge}), while in the limit of large $N$, this relation scales as $z_N \propto N^\alpha$. Therefore, we look for the exponents $\alpha$ that provide the best fits of the intrusion dynamics and we plot them in the inset of Fig.~\ref{fig:zj} as a function of the ratio $d/D$. We observe that $\alpha \simeq 0.33$ when $d/D \lesssim 0.3$ and decreases towards zero for larger values of the ratio $d/D$.\\
\indent When the friction on the lateral walls is mobilized, the pressure in the material does not increase linearly with depth, but rather saturates exponentially towards a constant value \cite{Janssen1895}. This corresponds to a dynamical Janssen effect resulting from the mobilization of the friction and is already observed when moving a vertical wall upwards \cite{Bertho2003}. However, in practice, we do not observe any saturation of the force experienced by a rod plunging into a granular medium \cite{Peng2009,Seguin2018,texier2022successive}. The moving wall is the side of the impactor that goes downwards and thus mobilizes friction in the opposite direction. In this case, the usual expression of the Janssen force adopts a different form since there is no saturation: we therefore consider that the pressure in the material increases exponentially with depth. This exponential dependence was already proposed to rationalize the force experienced by the bottom wall of a narrow granular column as it pushes up the grains \cite{Ovarlez2001}. This dependency of the pressure with depth is reflected in the drag force that applies on the impactor. Indeed, a previous study revealed that the presence of lateral walls increases the drag experienced by a sphere intruding at low speeds into a granular tank \cite{Katsuragi2012}. To account for this effect up to the limit of extreme confinements, we suggest an exponential variation of the force with depth
\begin{equation}
    m\frac{dv}{dt}= - K_z\, \rho gd \lambda^2 \exp\left(\frac{z}{\lambda}\right),
\end{equation}
where $\lambda$ is the characteristic depth of the evolution of the pressure profile. Following the same method as previously, the solution of this equation is:
\begin{equation}
    v^2(z) - v_i^2= - \frac{2 \,K_z\, \rho gd \lambda^3}{m} \left[ \exp\left( \frac{z}{\lambda}\right) - \exp\left( \frac{z_{N-1}}{\lambda}\right)\right].
\end{equation}
Looking for $z_N$ where $v(z_N)=0$, and using the characteristic length $L_0$ and the characteristic time $T_0$ introduced in the unbounded case, we obtain:
\begin{equation}
     \exp\left(\frac{z_N}{\lambda}\right) - \exp\left(\frac{z_{N-1}}{\lambda}\right)=\frac{v_i^2 L_0 T_0^2}{\lambda^3}.
\end{equation}
By iteration, we obtain a prediction for the intrusion dynamics in the presence of lateral confinement:
\begin{equation}
    \frac{z_N}{\lambda} =\ln\left[ \exp\left(\frac{z_0}{\lambda}\right) + \frac{v_i^2 L_0 T_0^2}{\lambda^3} N\right].
    \label{eq:intrusion_log}
\end{equation}
In the limit of large depths ($z_N \gg z_0 $), this prediction scales as $z_N \propto \ln N$ and thus has the same scaling as the observations made for $d/D \gtrsim 0.3$ (Fig.~\ref{fig:zj}). Since a power law of the form $z_N \propto N^\alpha$ is always larger than $\ln N$ at large $N$, this explains why the fits of the intrusion dynamics with Eq.~(\ref{eq:merge_alpha}) have an exponent $\alpha$ that decreases towards zero when the effect of the confinement dominates.

\section{Conclusion}
\label{conclusion}
In this study, we analyze the different intrusion regimes of a cylindrical impactor that penetrates a granular medium through $N$ successive impacts. In an unbounded medium, the dynamics of the impactor during the first impacts is ruled by a balance between the force resulting from the impact of the internal mass and the inertial resistance of the granular medium due to collision dissipation ($\propto v^2$), and leads to an intrusion law of the form $z_N \propto N$. For larger impact numbers, as the impactor moves deeper in the grains, the input energy is balanced by the quasi-static force applied by the granular medium on its lateral surface ($\propto z^2$). Thus, the intrusion dynamics becomes sublinear as $z_N \propto N^{1/3}$. We develop a theoretical framework that accounts for the forces applied by the grains on the impactor. The scaling laws predicted by the model are in agreement with the experimental measurements. The existence of these two regimes of penetration dynamics reveals a crossover depth $z_c \sim v_i(d/g)^{1/2}$ which depends on the geometry of the object and the impact velocity, also well captured by the model.

In a confined granular medium, the penetration dynamics is modified as the granular forces are modulated by the friction at the walls. The more confined the environment, the more difficult is the penetration by successive impacts since the resistant force increases. Consequently, the intrusion dynamics transits towards a logarithmic trend, $z_N \propto \ln N$ in the case of extreme confinement. We extend the theoretical framework that accounts for the forces applied by the granular medium on the impactor in this case.

The agreement between the predictions of this model and our measurements proves that we now have a reliable understanding of the granular forces applying on an impactor to predict its dynamics when its submitted to complex solicitations such as a series of impacts. This understanding offers promising opportunities to optimize the design of autonomous digging systems based on this principle and dedicated to explore any granular soils. In future studies, we can explore the effect of cohesion of the granular medium on the intrusion dynamics of the impactor to get closer to practical situations.

\acknowledgments
We are grateful to J.~Amarni, A.~Aubertin, L.~Auffray, C. Manquest, and R.~Pidoux for their contribution to the development of the experimental setup. We thank Hugo Gerard for preliminary experiments. This work was supported by ANR PIA funding: Grant No. ANR-20-IDEES-0002.

\bibliography{biblioJackhammer}

\end{document}